\documentclass[
10pt, 
a4paper, 
oneside]{article}
\pdfoutput=1

\usepackage[
nochapters, 
beramono, 
eulermath,
pdfspacing, 
dottedtoc 
]{classicthesis} 

\usepackage{arsclassica} 
\usepackage{helvet}
\usepackage[T1]{fontenc} 

\usepackage[utf8]{inputenc} 

\usepackage{graphicx} 
\graphicspath{{Figures/}} 

\usepackage{enumitem} 

\usepackage{lipsum} 

\usepackage{subfig} 

\usepackage{amsmath,amssymb,amsthm} 

\usepackage{varioref} 

\theoremstyle{definition} 

\theoremstyle{plain} 

\theoremstyle{remark} 

\hypersetup{
colorlinks=true, breaklinks=true, bookmarks=true,bookmarksnumbered,
urlcolor=webbrown, linkcolor=RoyalBlue, citecolor=webgreen, 
pdftitle={}, 
pdfauthor={\textcopyright}, 
pdfsubject={}, 
pdfkeywords={}, 
pdfcreator={pdfLaTeX}, 
pdfproducer={LaTeX with hyperref and ClassicThesis} 
} 

\usepackage[margin=1.2in]{geometry}

\usepackage{amsmath,amssymb}
\usepackage[british]{babel}
\usepackage{changepage}
\usepackage{xr}
\makeatletter
\newcommand*{\addFileDependency}[1]{
\typeout{(#1)}
\@addtofilelist{#1}
\IfFileExists{#1}{}{\typeout{No file #1.}}
}\makeatother

\usepackage{textcomp,marvosym}

\usepackage{cite}

\usepackage{nameref,hyperref}

\usepackage[right]{lineno}

\usepackage{microtype}
\DisableLigatures[f]{encoding = *, family = * }

\usepackage{setspace}

\usepackage{array}
\usepackage{caption}
\usepackage[affil-it]{authblk}

\begin{document}
\title{The effect of data augmentation and 3D-CNN depth 
on Alzheimer's Disease detection}

\author[1,2]{Rosanna Turrisi%
  \thanks{Corresponding author. Via Dodecaneso 35, 16146 Genova, Italy. Electronic address: \texttt{rosanna.turrisi@edu.unige.it}; Telephone: +39 010 353 6602}}
\affil[1]{Department of Informatics, Bioengineering, Robotics and System Engineering (DIBRIS), University of Genoa, Genoa, Italy}
\affil[2]{Machine Learning Genoa (MaLGa) Center, Genoa, Italy, University of Genoa, Genoa, Italy}

\author[1,2]{Alessandro Verri%
 }

\author[1,2]{Annalisa Barla%
 }
\author[ ]{for the Alzheimer's Disease Neuroimaging Initiative\thanks{Membership of the Alzheimer’s Disease Neuroimaging Initiative is provided in the Acknowledgments}
}
\date{}

\maketitle




\renewcommand{\sectionmark}[1]{\markright{\spacedlowsmallcaps{#1}}} 
\lehead{\mbox{\llap{\small\thepage\kern1em\color{halfgray} \vline}\color{halfgray}\hspace{0.5em}\rightmark\hfil}} 

\pagestyle{scrheadings} 



\setcounter{tocdepth}{2} 





\section*{Abstract} 
{\small
\paragraph{Background and Objectives}
Machine Learning (ML) has emerged as a promising approach in healthcare, outperforming traditional statistical techniques. However, to establish ML as a reliable tool in clinical practice, adherence to best practices regarding data handling, experimental design, and model evaluation is crucial. This work summarizes and strictly observes such practices to ensure reproducible and reliable ML. Specifically, we focus on Alzheimer's Disease (AD) detection, which serves as a paradigmatic example of challenging problem in healthcare. We investigate the impact of different data augmentation techniques and model complexity on the overall performance.

\paragraph{Methods}
We consider Magnetic Resonance Imaging (MRI) data from the Alzheimer's Disease Neuro\-imaging Initiative (ADNI) to address a classification problem employing 3D Convolutional Neural Network (CNN).  
The experiments are designed to compensate for data scarcity and initial random parameters by utilizing cross-validation and multiple training trials. Within this framework, we train 15 predictive models, considering three different data augmentation strategies and five distinct 3D CNN architectures, each varying in the number of convolutional layers. 
Specifically, the augmentation strategies are based on affine transformations, such as {\em zoom}, {\em shift}, and {\em rotation}, applied concurrently or separately.

\paragraph{Results}
The combined effect of data augmentation and model complexity leads to a variation in prediction performance up to 10\% of accuracy. When affine transformation are applied separately, the model is more accurate, independently from the adopted architecture. For all strategies, the model accuracy followed a concave behavior at increasing number of convolutional layers, peaking at an intermediate value of layers. 
The best model (\textbf{8 CL}, (B)) is the most stable across cross-validation folds and training trials, reaching excellent performance both on the testing set and on an external test set.   
\paragraph{Conclusions}
Our results emphasize several important insights in ML for AD diagnosis. Firstly, we demonstrate that the choice of data augmentation strategy plays a significant role in improving the performance of the models. Secondly, we highlight the importance of investigating the depth of the model architecture, as it has a measurable impact on the final performance. Lastly, our findings underscore the necessity for adhering to rigorous experimental practices in the field of ML applied to healthcare. 

 \paragraph{Keywords} Deep Learning, Alzheimer's Disease, Data Augmentation, Model Depth, Reproducibility









\section{Introduction}
Advanced Machine Learning (ML) techniques have proven to be highly effective in  healthcare applications, such as cancer detection and prognosis \cite{cruz2006applications,sajda2006machine,KOUROU20158, shen2019deep, chaunzwa2021deep} and heart diseases prediction \cite{mohan2019effective, palaniappan2008intelligent}. 
However, it is still premature to assert that ML has been widely accepted as standard in clinical practice. For instance, in \cite{roberts2021common} the authors reviewed thousands of papers on the use of ML to detect COVID-19, 
revealing that none of them achieved the required level of robustness and reproducibility necessary for their use in the medical field.
The ML community is rightly taking action to solve this issue, by establishing best practices 
\cite{heil2021reproducibility,pineau2021improving, beam2020challenges}
that meet the essential criteria of the scientific method \cite{ioannidis2005, stupple2019reproducibility} for producing high-quality publications and defining new medical protocols.

 \subsection*{Our contribution}%
To begin, we summarize the general guidelines for reproducible ML pertaining to two key aspects: \textit{data} and \textit{experimental design and model assessment}.
    
    \paragraph{Data}
    \begin{itemize}
    
     \item Data collection/selection should align with the scientific problem at hand, avoiding bias and information leakage (e.g., utilizing cross-sectional data for diagnostic confirmation or longitudinal data for prognostic purposes)\cite{jm2018data}. 
    
     \item Data quality should be assessed by identifying missing values and inconsistencies, and improved by applying appropriate imputation and cleaning methods\cite{lin2020missing}.    
    
     \item Data harmonization can be used to compensate for heterogeneous data from 
    different acquisition techniques\cite{kourou2018cohort}. 

     \item Data augmentation can be employed as a solution for small sample size or unbalanced samples per class, a common case in the biomedical field.
    
    
    \item The whole data handling process should be described in details in order to ensure reproducibility.
    \end{itemize}

    \paragraph{Experimental design and model assessment}

    \begin{itemize}

    \item  The versioned code used for conducting the experiments should be publicly shared to ensure transparency and reproducibility.
    
    \item Every decision in the design of the predictive model should be justified, with recognition of uncontrollable factors. \cite{haibe2020transparency}. 
    
    \item Details about the samples used in the the training/testing split should be diclosed to guarantee benchmarking.
 
    \item A well-designed experiment should avoid assessing results on a non-representa\-tive testing set. To this aim, resampling strategies \cite{batista2004study} such as k-fold cross-validation or boosting can be utilized to comprehensively assess the model's performance. 
    Further, models based on random weights initialization should be repeated for different trials in order to assess their stability.
    
    \item The performance metrics should be chosen according to the specific scientific objectives of the study \cite{sokolova2009systematic,chicco2020advantages}.
    
        
    \item Testing the model on external datasets is ideal to evaluate its generalization properties \cite{basaia2019automated}.
    
        \end{itemize}

    
Following these criteria, we took Alzheimer's Disease (AD) as a prominent example of complex disorder and we proposed a Deep Learning (DL) experiment to investigate the impact of data augmentation and model depth in a classification setting.  We addressed the problem of discriminating AD subjects from CN individuals by using low-resolution (1.5) MRI scans. We adopted 
a 3D-Convolutional Neural Network (CNN) \cite{lecun1995}, eliminating the need for feature engineering processes like ROI selection \cite{long2017prediction}. This setup is fairly ambitious due to the vast number of possible architectures and training strategies, but it eliminates the need of domain experts supervision.



A total of 15 DL models are compared, showing significant variations in prediction accuracy up to 10\%. One augmentation strategy consistently outperforms the others. Model accuracy varies as a concave  curve at increasing model depth values, peaking at an intermediate numbers of layers. 
The best model showed excellent accuracy on the testing set and good properties of generalization to an external dataset. 
It is worth noting that the proposed approach can be readily extended to other similar contexts beyond AD.\\

The paper is structured as follows. The Background section introduces the Alzheimer’s Disease Neuroimaging Initiative (ADNI) dataset as the standard for AD research and mentions relevant state-of-the art ML papers. The Materials and Methods section details data, methods, and experimental setup, including challenges and choices made. The Results section compares augmentation strategies and architectures. The Discussion section relates findings to state-of-the-art studies, and the Conclusion section illustrates future perspectives.

  \section{Background}
Many recent studies exploited ML methods to elucidate AD pathophysiological processes,  employing data from ADNI \cite{mueller, squillario2020telescope, Osipowicz2021, briones2012data, stokes2014application, Jiang2014, basaia2019automated,ZHANG2011856, Lian2020, venugopalan2021multimodal, counts2017biomarkers}. 
ADNI comprises heterogeneous datasets collected during different temporal phases (ADNI1, ADNI/GO, ADNI2, and ADNI3), each characterized by varying MRI acquisition protocols, see Fig. \ref{fig:ADNI_mri}. ADNI1 includes longitudinal acquisitions on 1.5T and 3T scanners with T1- and T2-weighted sequences; ADNI-GO/ADNI2 involves imaging data acquired at 3T with similar T1-weighted parameters to ADNI1; ADNI3 exclusively utilizes MRI obtained from 3T scanners.
  \begin{figure}[t!]
    \centering
    \includegraphics[width=0.9\textwidth]{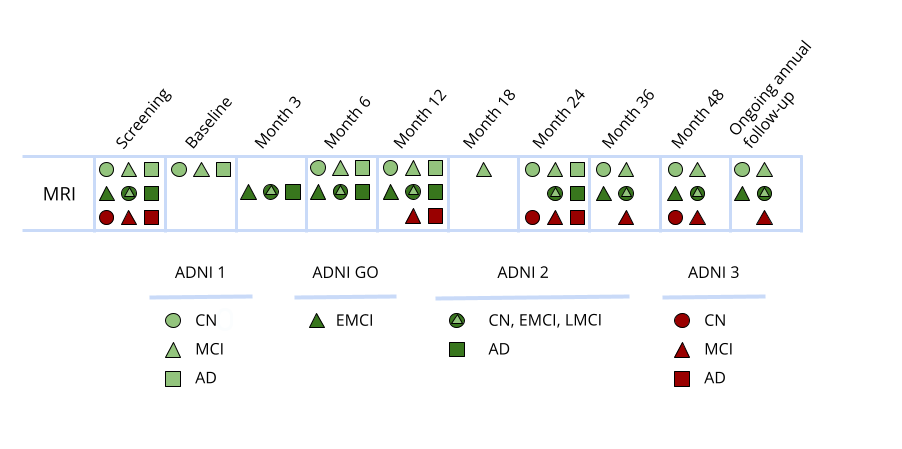}
    \caption{\textbf{MRI collection in ADNI dataset.} 
    Schema representing ADNI phases (ADNI1, ADNI GO, ADNI2, ADNI3). Different phases include a variable proportion of subjects:  circles represent CN subjects, triangles represent Mild Cognitive Impaired (MCI), early MCI (EMCI) or late MCI (LMCI) subjects and squares depict AD patients (picture inspired by the data samples schema in the ADNI website).}
    \label{fig:ADNI_mri}
\end{figure}


ADNI heterogeneity allowed for many different experimental setups in literature, with results depending on sample size (ranging from hundreds \cite{Liu2014, ALINSAIF2021104879, long2017prediction, korolev2017} to thousands  \cite{salehi2020CNN, basaia2019automated}), images resolution or sequence type. However, this flexibility and the lack of a universally recognized benchmark hampered a fair comparison among models' performance. Likewise, the absence of a standardized protocol for data handling, including dataset splitting and pre-processing, prevents the development of models transferable to the clinical practice. Despite having potential implications for future clinical applications, published results in this context often deviate from established principles of scientific methodology. In the remainder of this section, we outline significant findings in MRI-based classification on ADNI and discuss their experimental approach in relation to the criteria outlined in the previous section. 

In \cite{salehi2020CNN}, authors propose a 2D-CNN model to discriminate among CN, Mild Cognitive Impaired (MCI), and AD subjects reaching an accuracy of 99.3$\%$. Despite their model provides excellent results, the performed pipeline does not satisfy many data and experimental design criteria.
Authors state they adopt ADNI3 MRI for a total number of 6625 images. 
The Data Acquisition section reports that a total number of 7635 images is used in the work. In the same section, authors illustrates the data acquisition procedure in which they download 1290 MRI images from ADNI1 Annual 2 Yr 3T and ADNI1 Baseline 3T. This makes unclear which data has been selected for the experiments. 
Further, ADNI1 Baseline and ADNI1 Annual 2 Yr contain MRI exams from the same subjects at the baseline and after two years. We assume that, in two years, the MRI scan will not considerably change both for CN and for some stable patients. As a consequence, the model may have been trained and tested on very similar data. This clearly increases the performance results in terms of accuracy which, however, does not correspond to a real model improvement. We point out that longitudinal acquisitions at different pathological stages should be used only when modeling disease progression and outcome over time. Moreover, although the experimental design and the model are well described, the performance is only evaluated in terms of accuracy, disregarding other important measures, and on one trial only, discarding the weight robustness assessment. Finally, training/testing identifiers and the Python code are not publicly available. 

In \cite{korolev2017}, two different 3D-CNN architectures are proposed for performing AD/CN binary classification task, as well as other related tasks (e.g., AD vs early MCI (EMCI) subjects). The experiments are carried out by running 5-fold cross validation. The best architecture reaches 80$\%$ of average accuracy in discriminating between AD and CN. The ROC AUC is also reported. Both data and code are available on GitHub. Nonetheless, the proposed pipeline neglects some important aspects. Authors claim that in order to prevent information leakage they only select the first images for each subject within the ADNI dataset, for a total number of 231 images. However, it is not clear if they mixed data from different ADNI phases which would correspond to using heterogeneous images. Further, image processing consists of cropping images to $110 \times 110 \times 110$ size, an arbitrary choice that is not discussed. 
All experiments are run only once, without assessing the weight robustness. 

In \cite{basaia2019automated}, a CNN model takes 3T MRI exams from ADNI Baseline as input to perform a AD binary classification. Despite the promising result of 99.2$\%$ accuracy in differentiating between AD and CN, the model is tested only on a set of 65 samples, which may not be large enough to be representative.  

In \cite{ghaffari2022deep} the authors investigate the use of three popular pre-trained CNN models and their fine-tuning on 3T T1-weighted MRI from ADNI. Two tasks are performed: i) binary classification between CN and diseased subjects (including progressive MCI (pMCI), stable MCI (sMCI) and AD) ; ii) multi-class classification among CN, sMCI, pMCI, AD. Some essential details are missing on the preprocessing procedure, including how 2D images are obtained from volumetric MRI and how many synthetic samples are obtained by data augmentation. 
Results show that transfer learning always improves the classification performance on both tasks. AUC curve is also reported. However, the reported results may be unstable or not fully reliable as authors do not adopt any resampling strategy and the testing set only contains 32 samples. 
Remarkably, the authors also test their models on two external datasets reaching high accuracy. Nevertheless, both datasets have a very limited sample size (30 and 60 samples, respectively).


\section{Materials and Methods}




\subsection{Data}
For our experiments, we adopt the ADNI dataset \cite{mueller} considering only the ADNI1 data collected during screening, which is the baseline exam. This includes 550 1.5T MRI exams from 307 CN subjects and 243 AD patients. We use an additional set of 3T MRI exams to test the best model in a {\em domain shift} setting \cite{buchanan2021comparison}.
All data was preprocessed by ADNI experts (more information in \ref{appendix:data}). 

\subsubsection{Data processing}\label{processing}
We recall that MRI exams are three-dimensional data describing the structure of the brain. 
Fig. \ref{fig:2Dproj} displays the 2D projection of brain images captured from a CN subject (first row) and an AD patient (second row) on the \textit{sagittal}, \textit{coronal}, and \textit{axial} planes.
As already noted, ADNI images were collected with different protocols and scanning systems, hence they are very heterogeneous in size, see Table \ref{table1}. To enable the use of ML methods, it is necessary to select a common volume size. This choice, often unexplained in literature, defines fundamentals characteristics of the pipeline, such as the amount of information contained in the image and the input space dimension, on which model choice and computational burden depend.

In our experiments, images are downsized to $96\times 96 \times 73$. 
The principle guided this choice derives from computational issues. We first reduced the image dimension, rescaling the image of 50$\%$ along all dimensions, and we then resized images to match the smallest one. An alternative strategy may be zero-padding to match the biggest image but this would increase  memory requirements. 
 Finally, intensity-normalization was applied omitting the zero intensity voxels from the calculation of the mean. This procedure allows to have homogeneous data with a fixed size. 

\begin{figure}[!t]
  \centering  \includegraphics[width=0.5\textwidth]{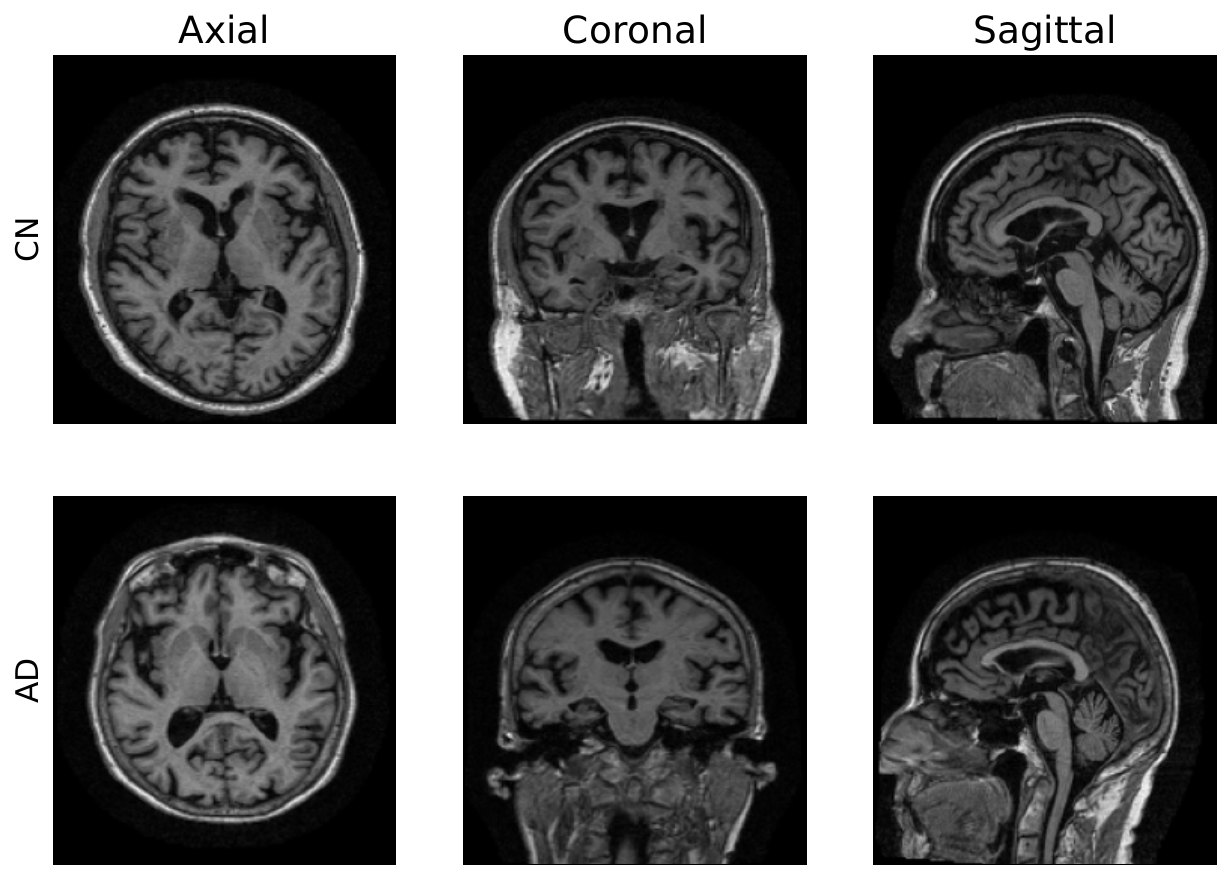}
    \captionof{figure}{\textbf{2D visualization of 3D MRI scans.} Axial, coronal and sagittal planes of two brain images from ADNI dataset.}
    \label{fig:2Dproj}
\end{figure}

\begin{table}[!t]
\centering
\caption{
\textbf{Baseline 1.5T ADNI1 dataset.} Number of CN and AD MRI scans grouped by size.}
\begin{tabular}{c|l|l|l}\small
\textbf{MRI size} & \textbf{CN} & \textbf{AD} & \textbf{Total} \\\hline
$256\times 256\times184$& 8& 8& 16\\\hline
$256\times 256\times170$& 40& 34& 74\\\hline
$256\times 256\times160$& 4& 0& 4\\\hline
$256\times 256\times166$& 97& 82& 179\\\hline
$256\times 256\times162$& 0& 1& 1\\\hline
$192\times 192\times160$& 117& 86& 203\\\hline
$256\times 256\times146$& 1& 0& 1\\\hline
$256\times 256\times161$& 2& 0& 2\\\hline
$256\times 256\times180$& 38& 32& 70\\
\end{tabular}
\label{table1}
\end{table}

\subsubsection{Data augmentation}

Data augmentation is a common procedure that simultaneously addresses data scarcity and creates a model invariant to a given set of transformations \cite{shorten2019survey}. Different augmentation strategies can result in varied training sets, affecting model performance and computational cost. In this study, the original set is augmented by applying separately or altogether \textit{zoom}, \textit{shift}, and \textit{rotation} transformations, as shown in  Fig. \ref{fig:data_augmentation} (see \ref{appendix:augmentation} for details). 
We devised the following strategies to compare the effect of different transformations and samples amount: 
\begin{itemize}
    \item \textbf{Strategy (A)}. To each image, we simultaneously apply all the transformations. The size of the augmented data will match the number of training samples $N$.
    \item \textbf{Strategy (B)}. To each image, we separately apply each transformation, generating three different distorted images. The size of the augmented data will be three times the number of training samples $3N$.
    \item \textbf{Strategy (C)}. To each image, we simultaneously apply all the transformations, as in strategy A. We repeat the process three times so that the number of augmented samples matches the one of strategy B ($3N$).
\end{itemize}

Therefore, strategies \textbf{(A)} and \textbf{(C)} rely on the same procedure, while strategies \textbf{(B)} and \textbf{(C)} generates the same number of samples. We remind that data augmentation is performed only on the training set, leaving validation and testing sets at the original sample size.

\begin{figure}[th!]
    \centering
    \includegraphics[width=0.4\textwidth]{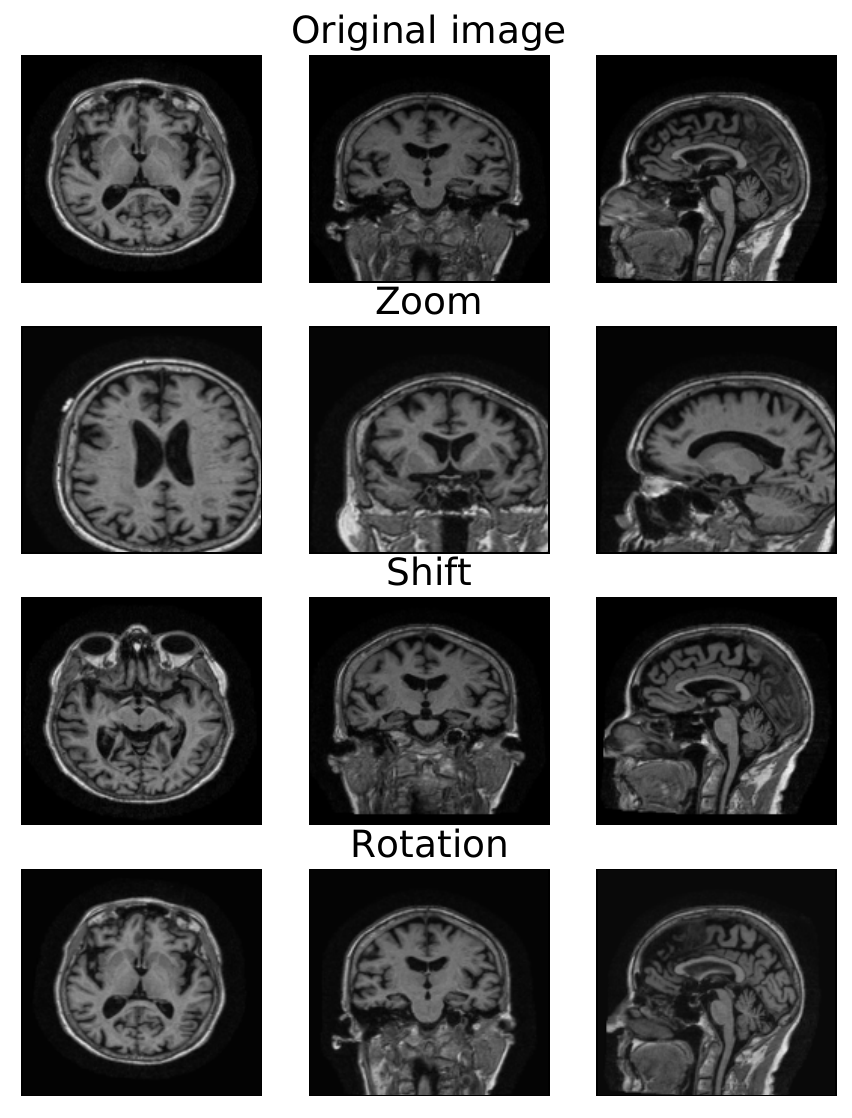}
    \caption{\textbf{Original and transformed MRI image.} 2D projections of the original MRI image (first row) and the augmented image obtained by applying {\em zoom} (second row), {\em shift} (third row), and {\em rotation} (last row) transformations.}
    \label{fig:data_augmentation}
\end{figure} 

\subsection{Experimental setup}

\subsubsection{Guide to the model choice}
Choosing the optimal DL model is not straightforward as the vast numbers of network and training parameters makes the brute-force approach unfeasible. Here we illustrate the model choices made a priori based on the issues posed by the addressed task.
\paragraph{Type of data}
Working with 3D images presents computational and memory challenges. As a solution, several studies in the literature adopt three 2D projections of the MRI. Nevertheless, this approach requires three separate models, leading to increased overall wall-clock time. Moreover, extracting features from the 2D projections may result in the loss of crucial volumetric information and a simplified representation of the studied phenomenon.
In this work, we adopted a 3D CNN that directly extracts volumetric features. 

\paragraph{Limited amount of data}
 To overcome the limited dataset size, we implemented the following strategies aimed at controlling model complexity and preventing overfitting: data augmentation; adding an $\ell _2$ penalty; and limiting the number of filters per layer. The latter method resulted in a substantial parameter reduction across the network. For instance, in a 2-layer CNN with 32 and 64 $3\times3\times3$ filters, reducing the number of filters to 8 and 16 (25\% of the initial values) leads to a significant reduction of 93\% in the number of learnable parameters (from 56256 to 3696).

\paragraph{Memory capacity} 3D models usually require a huge amount of memory capacity, that depends both on the input dimension and the model size. To reduce the required memory: i) we re-scaled the images to halve the data dimension;  ii) we chose a batch size that balances the memory cost while retaining a representative subset; iii) we balanced the number of filters and the batch size to reduce the computational burden of the activation layer. 


\subsubsection{Model details}
We report experiments on the CN/AD binary classification. 
A preliminary analysis performed with a standard training/validation/test split (75\%/15\% /10\%), denoted a very high variance due to the limited sample size of the testing set. For this reason, to guarantee a correct assessment of model performance and stability, we set up a stratified-K-fold cross-validation loop.
We set K=7, from Fold 0 to Fold 6 (training/validation/test, with a proportion of 70$\%$/15$\%$/15$\%$), that ensures having enough data for the learning phase. 
All folds are fully balanced, with the exception of Fold 6 which has an unbalanced ratio between AD and CN samples as the total amount of samples per class do not match exactly. 


\begin{figure}[t!]
    \centering
    \includegraphics[width=0.99\textwidth]{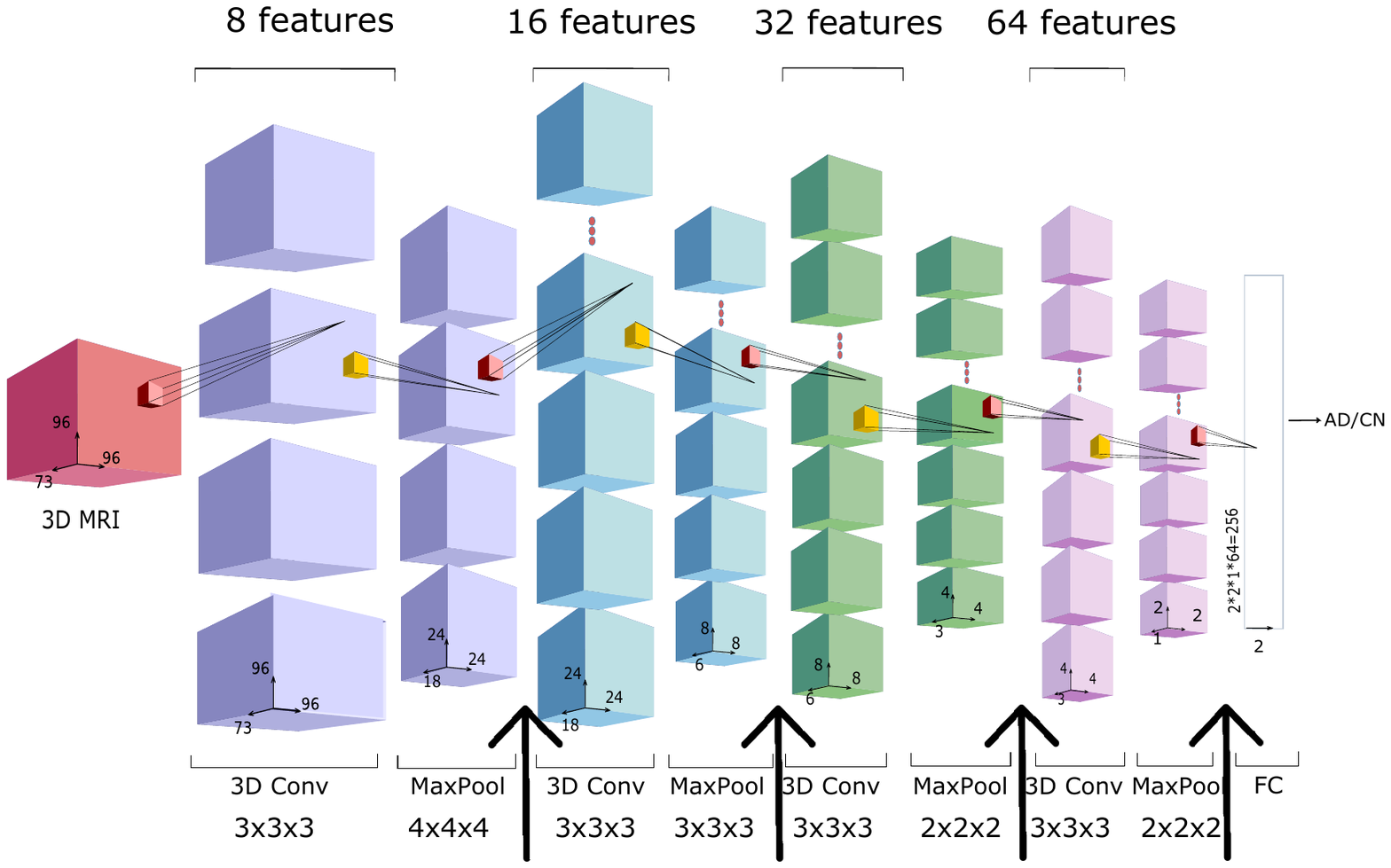}
    \caption{\textbf{3D-CNN Architecture.} Architecture of the \textbf{4 CL} baseline network, composed by four blocks of a convolutional and pooling layers, followed by a fully connected (FC) layer. The total number of features ($8*i$) in the $i$-th convolutional layer is marked above each layer, whereas the filter dimension is reported below. In the experiments, we consider other four extended versions of the baseline architecture duplicating the convolutional layer preceding the arrows.}
    \label{fig:cnn}
\end{figure} 

We adopted as baseline network an architecture with 4 Convolutional Layers (CL) followed by a fully-connected layer, as depicted in Fig. \ref{fig:cnn}. We will refer to this architecture as \textbf{4 CL} model. To investigate the optimal CNN depth, we inserted additional convolutional layers without pooling operations so that the number of layers is the only factor impacting in the model. Specifically, we added 2, 4, 6 and 8 convolutional layers in correspondence to the arrows of Fig. \ref{fig:cnn}. We refer to these models as \textbf{6 CL}, \textbf{8 CL}, \textbf{10 CL}, and \textbf{12 CL}. For instance, in the \textbf{10 CL} architecture 6 convolutional layers are added to the \textbf{4 CL} baseline: two layers are inserted in correspondence of the first and second rows, and one layer in correspondence of the third and fourth rows.
Additional details on network and training parameters can be found in \ref{appendix:experiments}.\\
In order to test model stability to initial random weights, each model has been run 10 times at fixed parameters. \\
All the experiments have been implemented using the Python programming language and performed on a Tesla K40c GPU. Samples identifiers and the Python code to reproduce the experiments are available on {\href{https://github.com/rturrisige/AD_classification}{GitHub}}.



\section{Results}
We compare 15 models obtained by combining different augmentation strategies with varying network depths, then we illustrate in detail the results of the best model. 
Results based on not-augmented data are not reported, as they were substantially worse than the ones obtained by using augmentation. 

\subsection{Architecture and augmentation choice}
Fig. \ref{fig:comparison_average} shows the accuracy on the validation set. As expected, Strategy (A) (in yellow) significantly underperforms the other augmentation types. This is due to the lower number of samples in the augmented data. Although strategies (B) (in green) and (C) (in fuchsia) generate the same amount of data, (B) outperforms (C) in all models, suggesting that applying the transformation separately significantly improves the CNN model. These outcomes hold independently from the adopted CNN architecture. 

\begin{figure}[t!]
    \centering
    \includegraphics[width=0.9\textwidth]{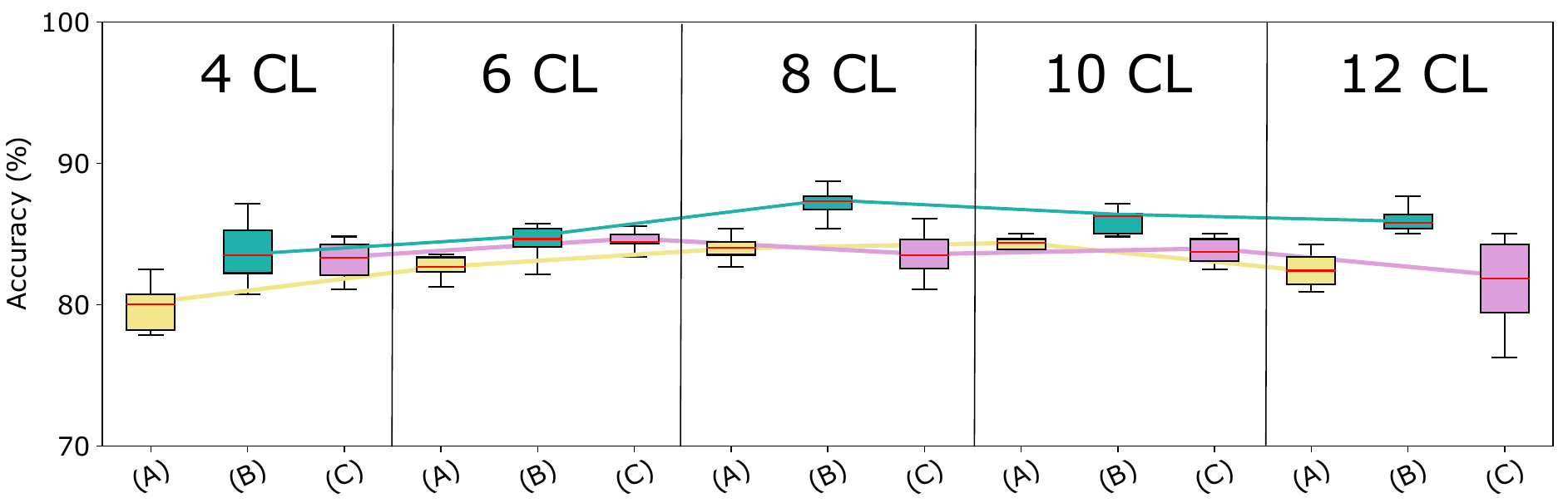}
    \caption{\textbf{Models accuracy at varying of architecture depth and augmentation strategies.} Comparison among the proposed CNN-based architectures with the three augmentation strategies, in terms of median accuracy on the validation set. The $y$-axis reports the model accuracy distribution on the 10 trials (\%) and the $x$-axis presents varying augmentation strategies (A), (B), and (C) in 5 blocks - one for each CNN architecture.}
    \label{fig:comparison_average}
\end{figure}

Moreover, the accuracy curves for all augmentation methods show a similar pattern: the best results are obtained for intermediate amounts of layers, while accuracy decreases for higher numbers of convolutional layers. The same behavior can been observed in Fig. \ref{fig:comparison_kfold} where we report for each cross-validation fold the distribution of accuracy in the 10 trials. The \textbf{8 CL} model with strategy (B) emerges as the best-performing combination, exhibiting also more stability compared to the other combinations.

\begin{figure}[t!]
    \centering
    \includegraphics[width=0.95\textwidth]{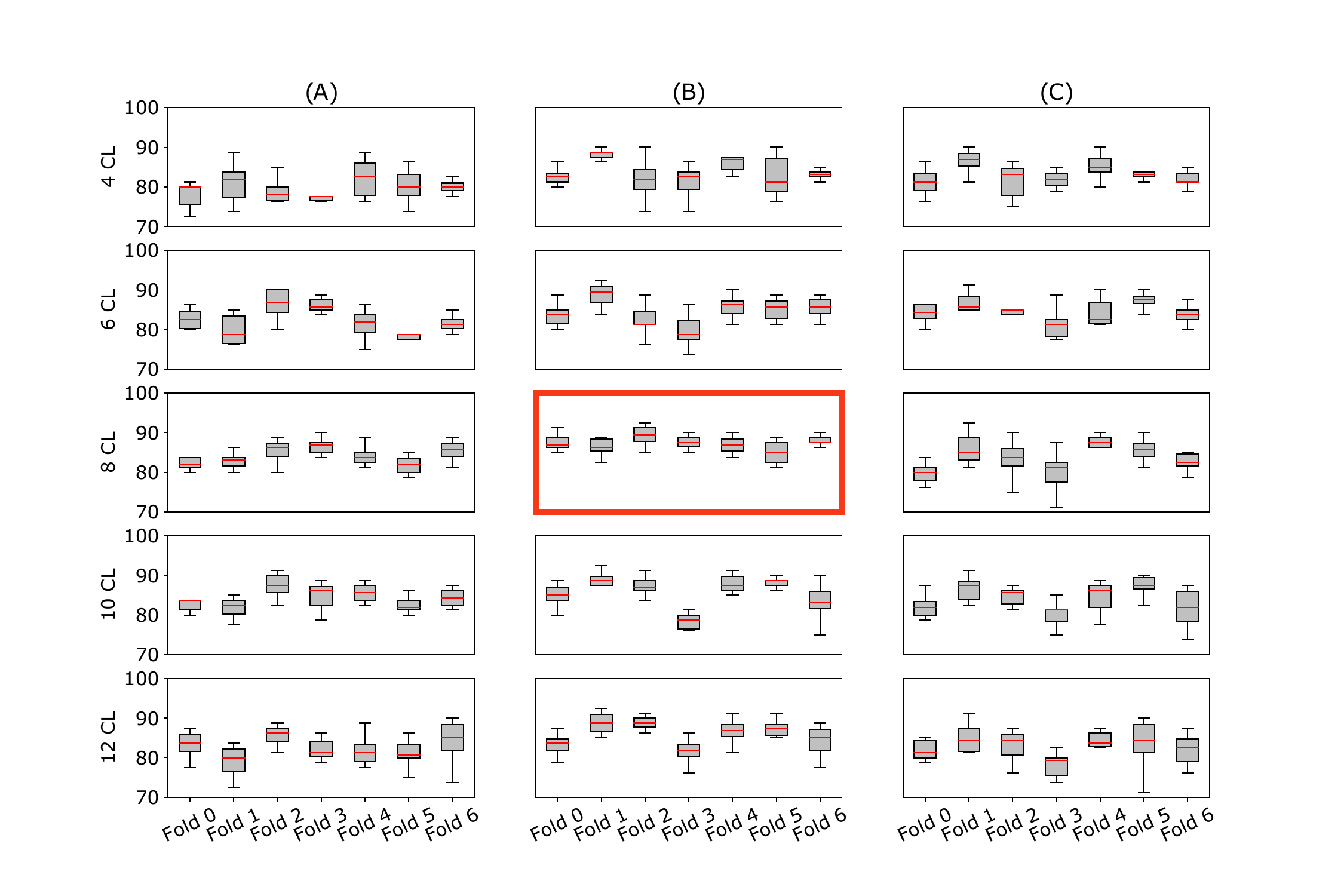}
    \caption{\textbf{Models performance and stability across folds.} {\em Small multiple} plot for the comparison of the validation accuracy for all architectures and augmentation strategies on all K-fold splits. On all subplots, the $y$-axis reports the model accuracy distribution on the 10 trials (\%) for each split ($x$-axis). Columns and rows display augmentation strategies and CNN architecture, respectively. The best combination (\textbf{8 CL}, (B)) is highlighted with a red border.}
    \label{fig:comparison_kfold}
\end{figure}

\subsection{Best model performance and insight}
The combination of a CNN with 8 convolutional layers and the (B) augmentation strategy (\textbf{8 CL}, (B)) turned out to be the best model, reaching an accuracy of $87.21\pm 0.88\%$ on the validation set and $81.95\pm 1.26\%$ on the testing set. 

\begin{figure}[t!]
    \centering
    \subfloat[\centering Complete evaluation of the model on CN and AD classes averaged over the 7 folds.]{{\includegraphics[width=0.6\textwidth]{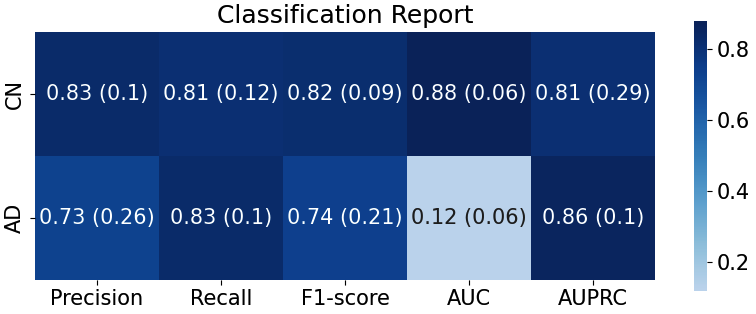} }}%
    \qquad
    \subfloat[\centering Confusion matrix of the classification results counted over the 7 folds.]{{\includegraphics[width=0.3\textwidth]{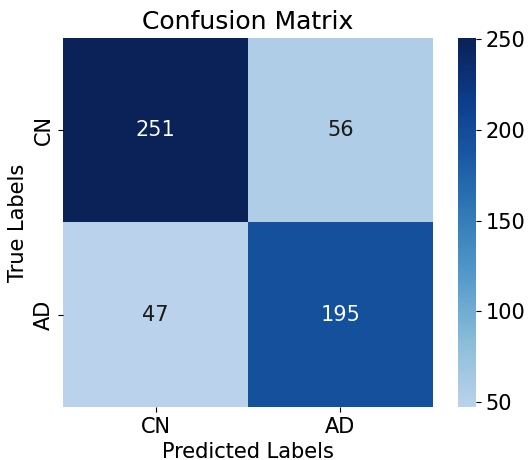} }}%
    \caption{Evaluation of the (\textbf{8 CL}, (B)) model on the testing set.}%
    \label{fig:modeleval}%
\end{figure}

A complete evaluation of this model is reported in Fig. \ref{fig:modeleval}: panel (a) resports mean and standard deviation for Precision, Recall, F1-score, AUC and AUCPRC of CN and AD classes over the 7 folds; panel (b) shows the Confusion matrix obtained by counting True Positive, True Negative, False Positive, and False Negative scores over the 7 folds. Fig. \ref{fig:layers} gives an insight on the layers behaviour and how they are learning the optimal model. Panel (a) displays the learned filters of every convolutional layer for one AD patient on the three considered median planes, i.e. {\em sagittal}, {\em coronal} and {\em axial}. 
It is clear that the filters capture more abstract features at increasing depth values. 
Panel (b) presents, for each convolutional layer, the layer outputs (\textit{embeddings}) of training and test samples projected on a two-dimensional plane through t-SNE \cite{van2008visualizing}. Both projections show that the embeddings are more clustered as the number of layers increases. \\

To further understand the properties and limits of the (\textbf{8 CL}, (B)) model, we assessed the effect of dropout, finding that it does not improve its performance (details in \ref{appendix:Dropout}). Also, we tested the model on an external dataset (described in \ref{appendix:3T_MRI}) of 3T MRI scans, obtaining an accuracy of 71\% and an AUC curve of 0.76 (a complete evaluation can be found in \ref{appendix:generalization}).

\begin{figure}[t!]
    \centering
    \subfloat[\centering Convolutional filters learned by the best model. ]{{\includegraphics[width=0.29\textwidth]{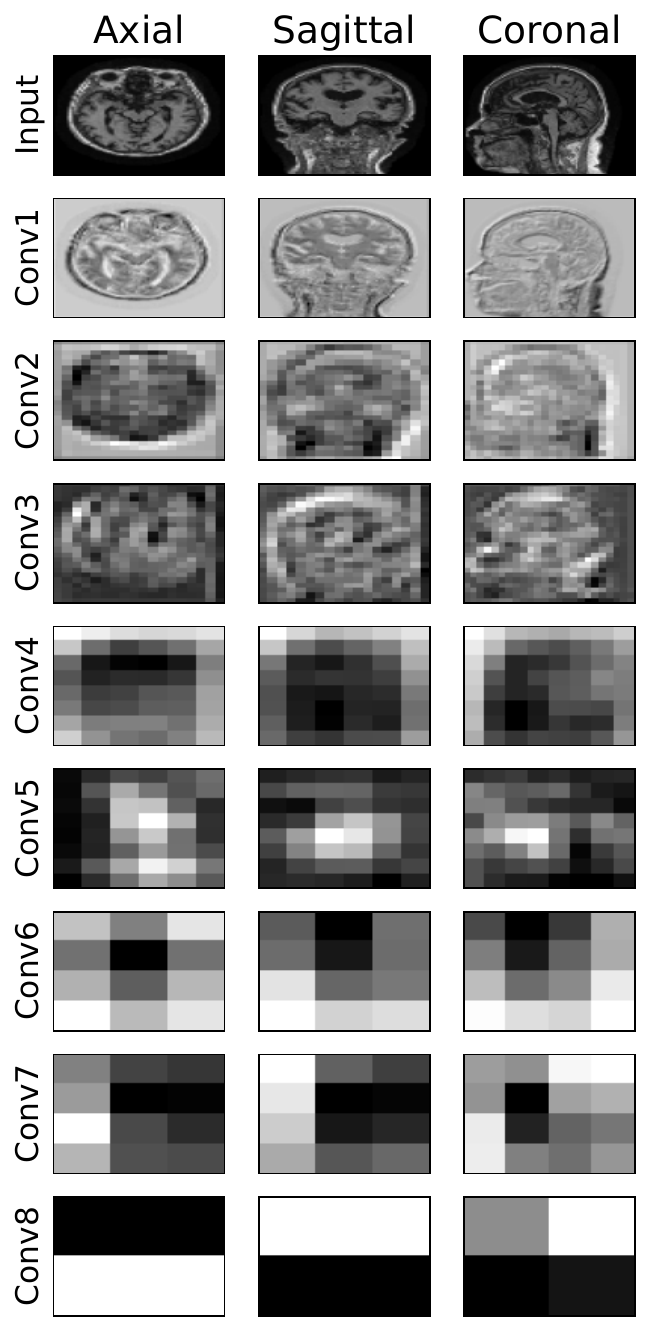} }}%
    \qquad
    \subfloat[\centering 2D projection of training and test embeddings learned by the best model.]{{\includegraphics[width=0.61\textwidth]{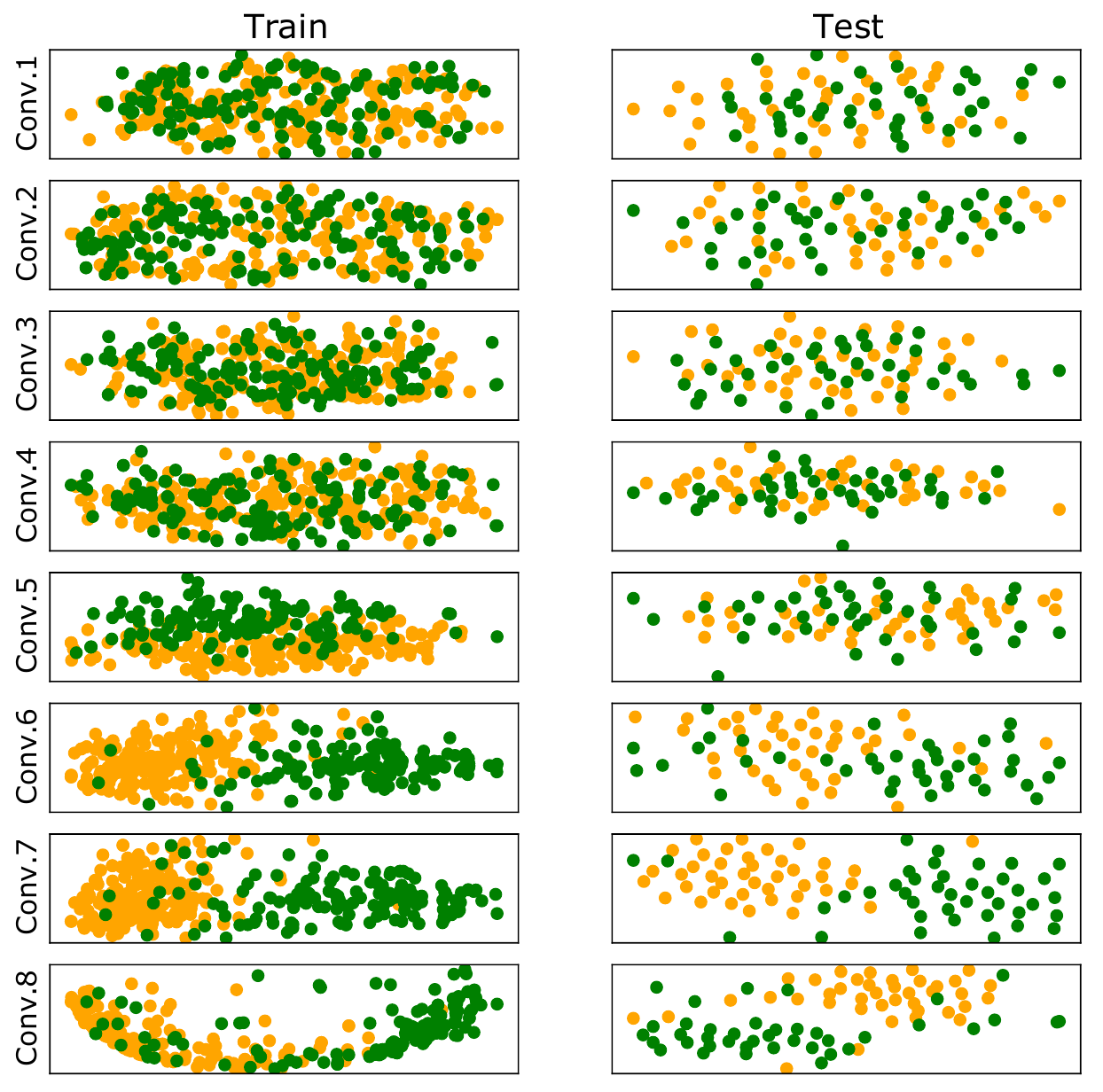} }}%
    \caption{(a) Illustration of the learned filters by the best model for one of the AD samples. Columns show filters for the three median planes and rows show the filters for the input (raw data) and the convolutional layers at increasing depth. (b) Training and test embeddings for each convolutional layer of the (\textbf{8 CL}, (B)) model projected by t-SNE. For increasing depth, AD (green) and CN (yellow)  samples are better clustered.}%
    \label{fig:layers}%
\end{figure}

\section{Discussion}
We analysed the impact of data augmentation strategies and number of convolutional layers in CNN models, considering a total of 15 combinations.
Independently from the adopted architecture, Strategy (B) always outperforms the others. As strategies (B) and (C) leverage the same amount of training samples, these results suggest that applying the affine transformations separately may be more effective than combining them simultaneously. 
We showed that models' performance can differ up to 10$\%$ average accuracy, highlighting the importance of correctly investigating the model depth and the set of data transformations. 
For all augmentation approaches, we found that the curve of the model accuracy at increasing depths tends to be a concave function reaching the maximum for an intermediate depth value. 
Although the widespread notion for which deeper neural networks better generalize in a general framework, this result is in line with other studies \cite{zhang2021understanding,vento2019traps} in which authors showed that smaller models perform better when only a limited amount of data is available, as they avoid overfitting. 

The best model we identified is the combination of a CNN with 8 convolutional layers and the (B) augmentation strategy (\textbf{8 CL}, (B)). 
The model accuracy in validation and testing is $87.21 \pm 0.88\%$ and $81.95\pm 1.26\%$, respectively, which is 4.2\% increase in accuracy with respect to (\textbf{4 CL}, (B)) model. Also, Fig. \ref{fig:comparison_kfold} shows how  (\textbf{8 CL}, (B)) is more stable than all other models with respect to both cross-validation folds and training trials. 
Although these results appear in line with current state-of-the-art studies, we argue that a true comparison is not completely feasible as other works employ different datasets and data types, the number of samples varies dramatically both in training and testing sets, experimental designs are very heterogeneous and, most importantly, performance is often assessed on one trial, without any variability estimation.
As additional evaluation, we tested the best model in a \textit{domain shift} context, i.e. on 3T MRI data, reaching 71\% of accuracy. We remark that this is a very challenging task as the image resolution deeply differs from the one in the training set.

\section{Conclusion}
This paper proposes an experimental pipeline for MRI-based binary classification of AD vs CN subjects, 
emphasizing key criteria for robustness and reproducibility.
The experiments have been conducted on a pre-processed subset of ADNI dataset that includes 1.5T MRI scans collected during the screening ADNI1. This selection ensures high data quality and harmonization, preventing any potential data leakage. The list of selected samples was made publicly available to enable benchmarking in further studies. Although the dataset is balanced, its sample size is limited. To address potential overfitting and ensure reliable results, resampling, data augmentation, and model complexity reduction strategies were employed. 

The first solution exploits K-fold cross-validation in order to provide a measure to model variability and robustness in terms of standard deviation.
The second approach augments the number of training samples by applying
affine transformations to the original image leading to a final sample size that depends on
how and how many transformations are applied.
The third strategy defines the model architecture in order to reduce the number of learnable parameters. 
In particular, the last two solutions require to select some parameters following empirical criteria which are often insufficiently discussed in literature.
We believe that if artificial intelligence aims at giving a real contribution in the daily clinical practice, ML methods should be designed and implemented following homogeneous and shared data acquisition protocols and benchmarks, standardized strategies for parameter selection and good practices that ensure robust and reproducible results. 

To the best of our knowledge this is the first work in the AD domain that digs into these experimental aspects and quantifies the impact on performance estimation. 
Future work will extend this analysis to other architectures, such as transformers \cite{jaderberg2015spatial}, additional affine transformations, different amounts of augmented samples, and, possibly, a multi-class classification setting that includes MCI subjects.

\section*{Acknowledgments}
{Rosanna Turrisi was supported by a research fellowship funded by the DECIPHER-ASL – Bando PRIN 2017 grant (2017SNW5MB - Ministry of University and Research, Italy).
Data used in preparation of this article were obtained from the Alzheimer's Disease Neuroimaging Initiative (ADNI) database (adni.loni.usc.edu). As such, the investigators within the ADNI contributed to the design and implementation of ADNI and/or provided data but did not participate in analysis or writing of this report. A complete listing of ADNI investigators can be found at: \href{http://adni.loni.usc.edu/wp\-content/ uploads/how\_to\_apply/ADNI\_Acknowledgement\_List.pdf}{ADNI Acknowledgment List.}
}


\renewcommand{\refname}{\spacedlowsmallcaps{References}} 

\bibliographystyle{unsrt}

\bibliography{main.bib} 


\end{document}



\renewcommand{\sectionmark}[1]{\markright{\spacedlowsmallcaps{#1}}} 
\lehead{\mbox{\llap{\small\thepage\kern1em\color{halfgray} \vline}\color{halfgray}\hspace{0.5em}\rightmark\hfil}} 

\pagestyle{scrheadings} 

\maketitle
\begin{center}
\textbf{\huge Appendix}
\end{center}

\appendix

\section{Data}\label{appendix:data}
\subsection{The ADNI dataset}
The ADNI cohort is a longitudinal multicenter study that aims at developing clinical, genetic and biomedical biomarkers for AD early detection. ADNI started collecting data since early $2000$s and experienced four different phases, including over 2000 subjects affected by different degrees of cognitive impairment. However, different data modalities are not available for all subjects and therefore some scientific questions are still out of reach due to data scarcity.\\

In this work, we considered a pre-processed subset of 550 1.5T T1-weighted MRI scans from ADNI1. This includes 307 CN subjects and 243 AD patients. \\
Further, to evaluate the model ability to generalize to new datasets in the context of \textit{domain shift}, we utilized 3T T1-weighted MRI scans as dataset for external validation. Specifically, this consists of 80 pre-processed images from ADNI1, including 47 healthy subjects and 33 patients with AD.

\subsection{Data pre-processing}
Experts by ADNI pre-processed the MRI exams in order to correct the image geometry distortion (by Gradwarp), the image intensity non-uniformity (by B1 calibration and N3). Finally, the images have been scaled for gradient drift using phantom data as detailed on the ADNI website. Note that not all these techniques have been simultaneously applied to each image, as the preprocessing procedure varies based on the acquisition system. It is worthwhile to mention that the ADNI screening folder contains images with a second type of scaling, referred to as  \textit{scaled$\_$2}, which we used in place of the \textit{scaled} when it is not available. The ADNI consortium did not report details on differences between the two scaling methods.

\subsection{Data augmentation}\label{appendix:augmentation}
In this work, we studied the effect of various augmentation strategies in which we differently combined three affine transformations, described in the following.
\begin{itemize}
    \item \textit{Zoom}. The in/out zoom is applied by randomly generating the zoom percentage, in a range from 0 to 20$\%$. 
    \item \textit{Shift}. The shift is differently generated for each image dimension, so that shift$<0.4$.
    \item \textit{Rotation}. The rotation is defined by randomly generating a rotation angle, between -5 and 5 degrees, for each image dimension. 
\end{itemize}

It is clear that there is not a golden rule for picking the optimal transformation parameters and that any value is somewhat arbitrary. Here we assume that brain image acquisitions differ one from another only for small variations. This assumption motivated our choices. 

\section{Experimental setup}\label{appendix:experiments}
We adopted as baseline network an architecture with 4 Convolutional Layers (CL) followed by a fully-connected layer (\textbf{4 CL} model).
The number of filters in the $i$-th layer was set to $8*i$. All convolutional layers have filters with a $3\times3\times3$ kernel. Padding is performed so that the original image and the feature map have the same size. We applied batch normalization to each convolutional layer. Successively, pooling was applied with decreasing size (i.e., $4\times4\times4$, $3\times3\times3$, $2\times2\times2$, $2\times2\times2$). The pooling size was chosen in order to decrease the layer size and, consequently, the computational cost. To investigate the optimal CNN depth, we inserted additional convolutional layers without pooling operations so that the number of layers is the only factor impacting in the model. 
Specifically, we 2, 4, 6 and 8 convolutional layers obtaining five models increasing number of layers 4, 6, 8, 10, 12.

Each model was trained using the Adam optimizer \cite{kingma2014adam} with a learning rate set to 0.001. We trained the network to minimize the cross-entropy loss function with $\ell _2$-penalty weighted by 0.01. We allowed a maximum number of 200 epochs using early stopping if the performance does not increase after 20 epochs (patience). The batch size was 50. The choice of the described parameters was guided by the the criteria discussed in the Materials and Methods section and an exploratory analysis on the smallest model. 

\begin{figure}[t!]
    \centering
    \includegraphics[width=0.6\textwidth]{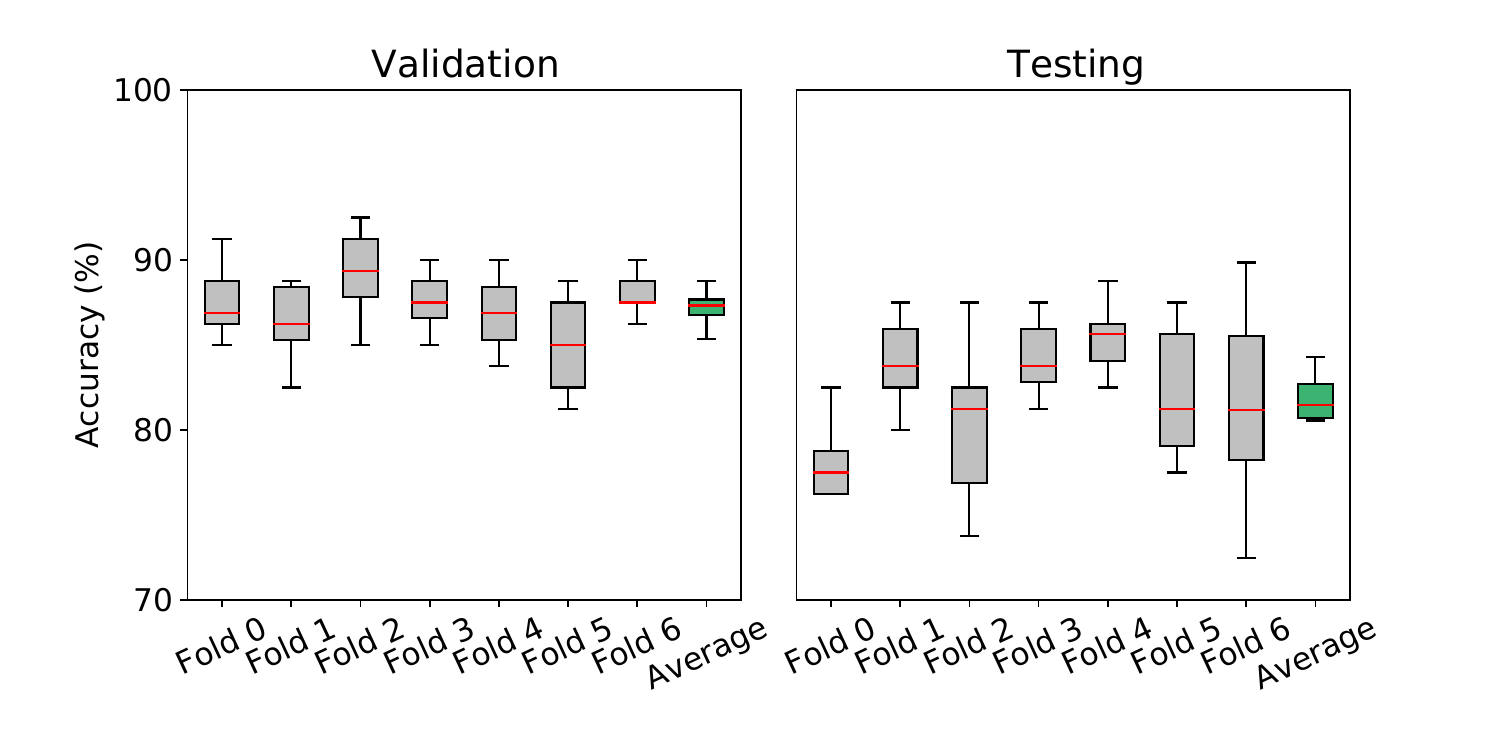}
    \caption{\textbf{Validation and test accuracy of best model.} Evaluation of (\textbf{8 CL}, (B)) model on validation and testing set in terms of percentage accuracy distribution on each fold (in silver) of the cross-validation and on average (in green), for all trials.}
    \label{fig:8CL_val_test_acc}

    \centering
    \includegraphics[width=0.6\textwidth]{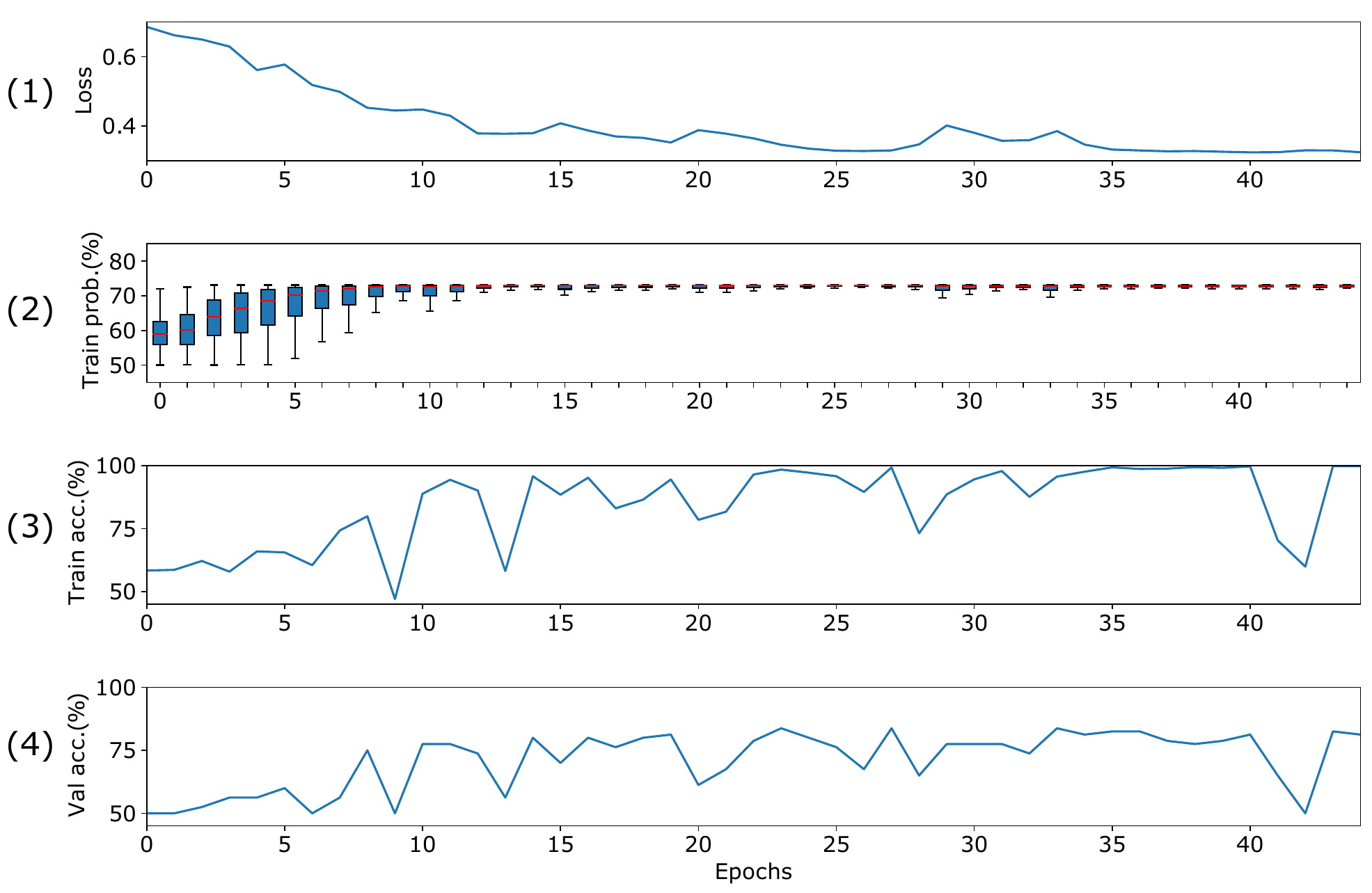}
    \caption{\textbf{Evaluation of the best model during training.} Four plots presenting, at evolving epochs, (1) the cross-entropy loss function values, (2) the probability distribution of training outputs, (3) training accuracy and (4) validation accuracy of the (\textbf{8 CL}, (B)) model.}
    \label{fig:baseline_train}
\end{figure}

\section{Results}

\subsection{Best model performance and insight}

Fig. \ref{fig:8CL_val_test_acc} shows the model performance distribution on validation and testing set for all trials. Note that performance on Fold 6 of the testing set is associated with a very high variability due to the unbalanced AD/CN ratio, as anticipated in the Methods and Materials section. To better describe the behavior of the best model, we explored how some key characteristics change with evolving epochs, see the plots in Fig. \ref{fig:baseline_train} .

In subplot 1, we can look at the cross-entropy loss function to assess proper model convergence as it decreases over time (epochs). 
In subplot 2, we can observe the training probability distribution of the predicted class, which is represented by the output of the last feed-forward layer. Here we verified that the learning process was not happening in an overfitting regime, as the training probability improves in terms of median and variability, but never reaching 100\%. 
Subplots 3 and 4 display the training and validation accuracy across the epochs: their behavior is comparable, confirming that the model shows very good generalization properties. We noticed that both accuracy curves do not show a step-wise behavior. Hence, in further studies, one could reduce the patience (currently, 20) in the early stopping criterion and stop the training when the performance does not increase after a few epochs (e.g. 5 epochs). This would significantly reduce the computational cost and the training time of the experiments.

\subsection{Ablation study on Dropout}\label{appendix:Dropout}
Table \ref{dropout_table} reports the classification accuracy on the validation and testing sets and the number of training epochs for the (\textbf{8 CL}, (B)) model in which dropout is applied with dropping probability ranging from 0 (i.e., no dropout) to 0.5. All the results are averaged on 10 trials. All models are comparable in terms of computational cost, whereas (\textbf{8 CL}, (B)) model without dropout slightly outperforms the others.

\begin{table}[!t]
\caption{Performance of (\textbf{8 CL}, (B)) model with and without dropout.}
\begin{center}
\begin{tabular}{c|c|c|c}
\textbf{Dropout} & \textbf{Validation accuracy} & \textbf{Testing accuracy} & \textbf{N. epochs} \\
\hline
0 &  $87.21\pm 0.88$ & $81.95\pm 1.26$ & 51\\
0.1 &  $85.64\pm 2.56$ & $80.21\pm 3.18$ & 51\\
0.25 & $86.09\pm 1.34$ & $80.57\pm 3.55$ & 51\\
0.5 & $86.02\pm 1.69$ & $80.96\pm 2.50$ & 50\\
\end{tabular}
\label{dropout_table}

\end{center}
\end{table}

\begin{figure}[!t]
    \centering
    \includegraphics[width=0.7\textwidth]{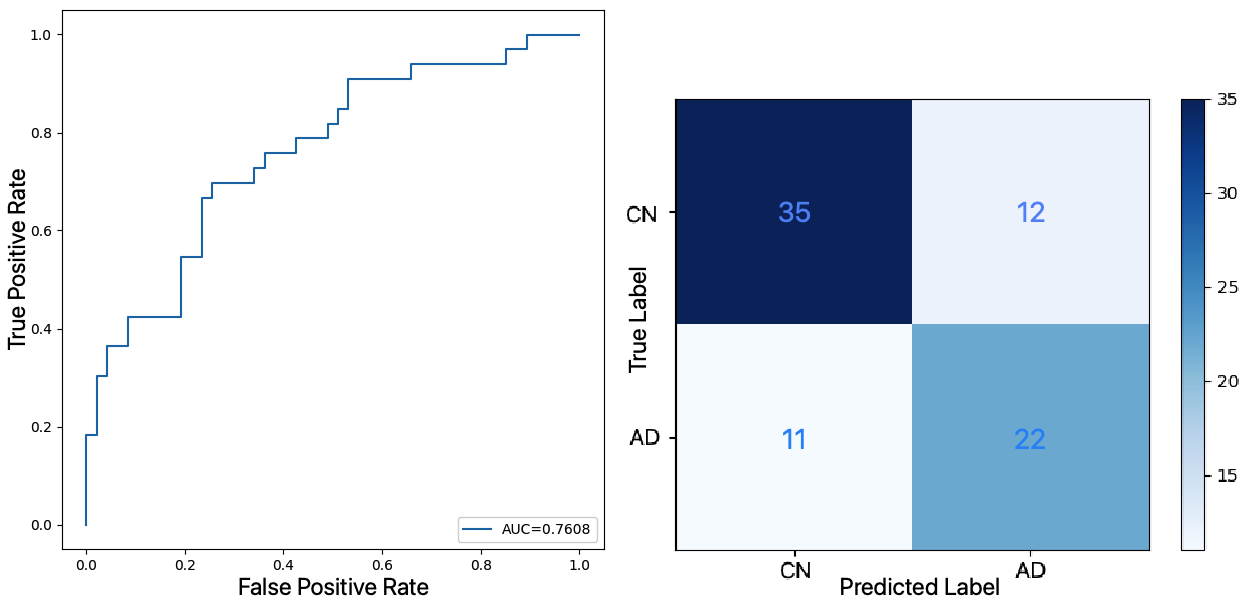}
    \caption{\textbf{Results on 3T MRI scans.} (left) AUC curve, (right) Confusion matrix.}
    \label{fig:generalization}
\end{figure}
\subsection{Generalization across image resolution}\label{appendix:generalization}
We evaluated the prediction ability of the (\textbf{8 CL}, (B)) model on unseen dataset that presents a shift domain. Specifically, we tested the generalization across image resolution. Results provided 71\% of accuracy and an AUC curve of 0.76. Fig \ref{fig:generalization} reports the AUC curve (left) and the the confusion matrix (right).

\renewcommand{\refname}{\spacedlowsmallcaps{References}} 

\bibliographystyle{unsrt}

\bibliography{main.bib} 